\documentclass[superscriptaddress,reprint,aps,prd]{revtex4-2}
\usepackage{amsmath}
\allowdisplaybreaks[4]
\usepackage{multirow}
\usepackage{amsfonts}
\usepackage{xcolor}
\usepackage{comment}
\usepackage{graphicx}
\usepackage{amsmath}
\usepackage{siunitx}
\usepackage{lineno}
\usepackage{float}
\usepackage{ulem}
\usepackage[
    pdfauthor={Teng Zhang},
    pdftitle={A Gravitational Wave Detector for Post-Merger Neutron Stars},
    colorlinks=true,
    linkcolor=black,
    urlcolor=blue,
    citecolor=blue
]{hyperref}
\DeclareSIUnit{\sqrthz}{\ensuremath{\sqrt{\text{\hertz}}}}

\begin{document}
\renewcommand\texteuro{FIXME}

\allowdisplaybreaks[4]
\title{A Gravitational Wave Detector for Post-Merger Neutron Stars: Beyond the Quantum Loss Limit of Michelson Fabry–P\'erot Interferometer}

\author{Teng Zhang}
\affiliation{School of Physics and Astronomy, and Institute for Gravitational Wave Astronomy, University of Birmingham, Edgbaston, Birmingham B15\,2TT, United Kingdom}
\affiliation{Department of Gravitational Waves and Fundamental Physics, Maastricht University, 6200 MD Maastricht, The Netherlands}
\affiliation{Nikhef, Science Park 105, 1098 XG Amsterdam, The Netherlands}

\author{Huan Yang}
\affiliation{Perimeter Institute for Theoretical Physics, Waterloo, Ontario N2L 2Y5, Canada}
\affiliation{University of Guelph, Guelph, Ontario N1G 2W1, Canada}

\author{Denis Martynov}
\affiliation{School of Physics and Astronomy, and Institute for Gravitational Wave Astronomy, University of Birmingham, Edgbaston, Birmingham B15\,2TT, United Kingdom}

\author{Patricia Schmidt}
\affiliation{School of Physics and Astronomy, and Institute for Gravitational Wave Astronomy, University of Birmingham, Edgbaston, Birmingham B15\,2TT, United Kingdom}

\author{Haixing Miao}
\affiliation{Department of Physics, Tsinghua University, Beijing, China}
\affiliation{Frontier Science Center for Quantum Information, Beijing, China}

\begin{abstract}
Advanced gravitational-wave detectors that have made
groundbreaking discoveries are Michelson interferometers with resonating optical
cavities as their arms. As light travels at a finite speed, these cavities are
optimal for enhancing signals at frequencies within the bandwidth, beyond which, however, a small amount of optical loss will
significantly impact the high-frequency signals. 
We find an elegant interferometer configuration with an ``L-resonator" as the core, significantly surpassing the loss-limited sensitivity of dual-recycled-Fabry–P\'erot-Michelson interferometers at high frequencies. Following this concept, we provide a broadband design of a 25\,km detector with outstanding sensitivity between 2\,kHz and 4\,kHz. We have performed Monte-Carlo population studies of binary neutron star mergers, given the most recent merger rate from the GWTC-3 catalog and several representative neutron star equations of state. We find that the new interferometer configuration significantly outperforms other third-generation detectors by a factor of 1.7 to 4 in the signal-to-noise ratio of the post-merger signal. Assuming a detection threshold with signal-to-noise ratio $>5$ and for the cases, we have explored, the new design is the only detector that robustly achieves a detection rate of the neutron star post-merger larger than one per year, with the expected rate  between $\mathcal{O}(1)$ and $\mathcal{O}(10)$ events per year.
\end{abstract}
\maketitle

\section{Introduction}
In 2015, the first direct detection of gravitational waves was made by the LIGO-Virgo collabration\,\cite{Abbott2016}. Since then, gravitational waves have become a new window for observing the universe and probing the unexplored territories in astrophysics, cosmology, and fundamental physics. Until now, more than 90 compact binary merger events have been confidently observed by the network of advanced detectors, including Virgo and KAGRA\,\cite{abbott2021gwtc}. 
Within these events, the detection of a binary neutron star coalescence, GW170817\,\cite{PhysRevLett.119.161101} followed shortly by a short gamma-ray burst, GRB 170817A\,\cite{2017_ApJ.848.L13_LVC} and the a series of joint observations of electromagnetic counterparts\,\cite{Abbott_2017} have had profound scientific impacts: This multi-messenger discovery confirmed that binary neutron star mergers are the origin of at least some short gamma-ray bursts and a production site for heavy elements via rapid neutron-capture\,\cite{Abbott_2017}. In the cosmology aspect, GW170817 has led to an independent measurement of the Hubble constant\,\cite{NatureHC}. It also provided unique access to probe the internal structure of neutron stars and their equation of state by constraining their tidal deformability\,\cite{PhysRevLett.119.161101,LIGOScientific:2018cki, LIGOScientific:2018hze, LIGOScientific:2019eut,Chatziioannou:2020pqz,Pan:2020tht}. With both gravitational wave and gamma ray measurements, new  constraints/bounds have been placed on the speed of gravitational waves and the violation of Lorentz invariance, in addition to a new test of the equivalence principle\,\cite{2017_ApJ.848.L13_LVC}.

Current gravitational wave detectors are only sensitive to the inspiral part of binary neutron star mergers, as shown in the analysis of GW170817.
The post-merger gravitational wave signal, which concentrates in the kilo-hertz band, encodes essential information to answer many important questions, \textit{e.g.} the origin(s) of heavy element nucleosynthesis, the engine of gamma-ray jets and the inner structure of the neutron star under extreme conditions. In particular, the merger and post-merger signals provide access to completely unexplored regimes in the Quantum Chromodynamics phase diagram beyond the reach of terrestrial collision experiment\,\cite{Schmidt_2017}, where the novel phase of matter may appear, \textit{e.g.} from hadron-quark phase transition\cite{PhysRevD.80.123009,PhysRevLett.122.061101,PhysRevLett.122.061102}.
\begin{figure}[t]
     \centering
     \includegraphics[width=1\linewidth]{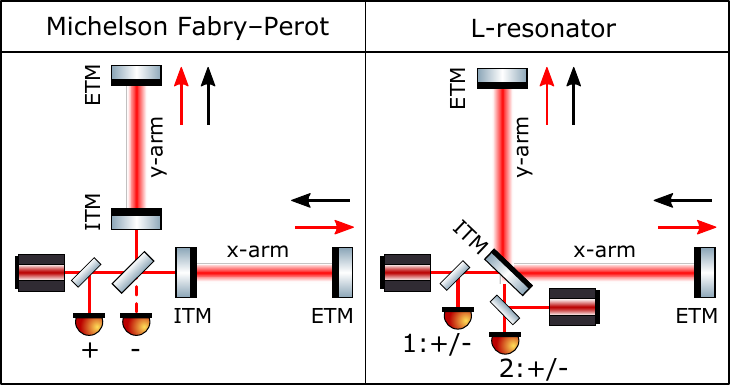}
     \caption{Schematic of FPMI interferometer and the "L-resonator". They are both sensitive to two degrees of freedom of motion: naming common,``+", mode indicated by red arrows, and differential,``-", mode (gravitational wave mode) indicated by black arrows.}\label{fig:Cavity}
\end{figure}

The successful performance of current gravitational wave detectors relies on their Michelson-type design. In addition to the canonical Michelson interferometer configuration, extra mirrors have been introduced in the arms to form Fabry–P\'erot cavities which boost both the optical power and the gravitational wave signals, giving so-called Fabry–P\'erot-Michelson (FPMI) interferometer\,\cite{drever1983laser}. With these modifications, the shot noise limited sensitivity is improved by orders of magnitude within the cavity bandwidth, typically from a few hertz to tens of hertz.

On the other hand, the binary neutron star post-merger signals are mainly between 2\,kHz and 4\,kHz which is beyond the optimal band of current gravitation wave detectors. To better explore neutron star physics, there are various ideas developed to improve the high-frequency sensitivity of modern and future detectors based on the dual-recycled-Fabry–P\'erot-Michelson (DRFPMI) interferometer, which includes a power recycling cavity at the bright port and a signal recycling cavity/signal extraction cavity (SEC) at the dark port to further enhance the arm cavity power and adjust the detector response\,\cite{mcclelland1995overview,PhysRevD.67.062002}. One straightforward approach is to make use of the adjustability of SEC, which forms a coupled system with the arm cavity. For example, the signal resonant frequency of the detector can be shifted to higher frequencies by constant SEC detuning\,\cite{PhysRevD.103.022002} or exploring the SEC-arm coupled cavity resonance\,\cite{PhysRevD.99.102004,NEMO,PhysRevD.102.122003}. More sophisticated quantum schemes, including the white light cavity\,\cite{Wicht1997,Zhou2015, PhysRevLett.115.211104,PhysRevD.99.102001,page2020gravitational,PhysRevD.103.122001} and the nonlinear optical parametric amplifier\,\cite{korobko2019quantum,Adya_2020,PhysRevD.106.L041101,li2020broadband,PhysRevD.106.082002}, aim at  broadening the effective bandwidth of the detector without sacrificing its peak sensitivity, hence overcoming the Mizuno limit\,\cite{MIZUNO1993273}. However, it is the optical loss that sets a universal and ultimate sensitivity limit of a quantum detector\,\cite{PhysRevX.9.011053}.
The schemes mentioned above for improving the high-frequency sensitivity are all severely constrained by the optical losses in the SEC, which directly attenuate the signal emerging from the arm cavity. Even worse, the SEC loss limited high-frequency sensitivity is independent of the arm length.
Several studies have been carried out to explore various techniques to saturate or overcome the SEC loss limit\,\cite{galaxies9010003,PhysRevD.104.122003}. It has been realised that a sloshing-type Sagnac configuration can beat the SEC loss limit by adding a filter cavity between two arms and thus shaping coupled cavity resonances in the absence of SEC. However, the filter cavity loss becomes the new limiting factor as another internal loss\,\cite{PhysRevD.104.122003}. 
Physically the intrinsic limit of high-frequency sensitivity comes from the decay of signal beyond the bandwidth of the single cavity where the signal is generated and circles around. 

In order to surpass the loss limit of gravitational wave detectors at high frequencies, the question becomes whether it is possible to resonate high-frequency signals in the arm cavity by itself. In the meanwhile, the laser carrier needs to resonate to maintain the high power. It motivates the idea of taking advantage of resonance at free spectral range away from the carrier frequency\,\cite{Zhang:2020qlh}. However, the detector's response to gravitational waves essentially degrades around the free spectral range frequency\,\cite{Rakhmanov_2008,Rakhmanov_2009,PhysRevD.96.084004}. For gravitational waves from the normal direction, the alteration in travel time for light cancels out if the duration of the round trip matches the period of the gravitational wave.

In this paper, we provide an elegant detector scheme satisfying all criteria. The SEC loss limited lower bound sensitivity of the new detector can be orders of magnitude better than DRFPMI interferometer at high frequencies. A conceptual design of the new detector, including all other losses, surpasses the quantum loss limit of currently proposed gravitational wave detectors several times. In Sec.~\ref{sec:Principle}, we introduce the principle of the core of the new detector, an ``L-resonator". In Sec.~\ref{sec:Quantum noise}, we present the complete interferometer and its quantum noise. 
In Sec.~\ref{sec:Conceptual}, we deliver the conceptual design of the new detector and model the noise budget. In Sec.~\ref{sec:Science}, we demonstrate the ability of the new detector to detect the post-merger signal of binary neutron star coalescence and dark-matter-induced neutron star collapse. 
\begin{figure}[t]
     \centering
     \includegraphics[width=1\linewidth]{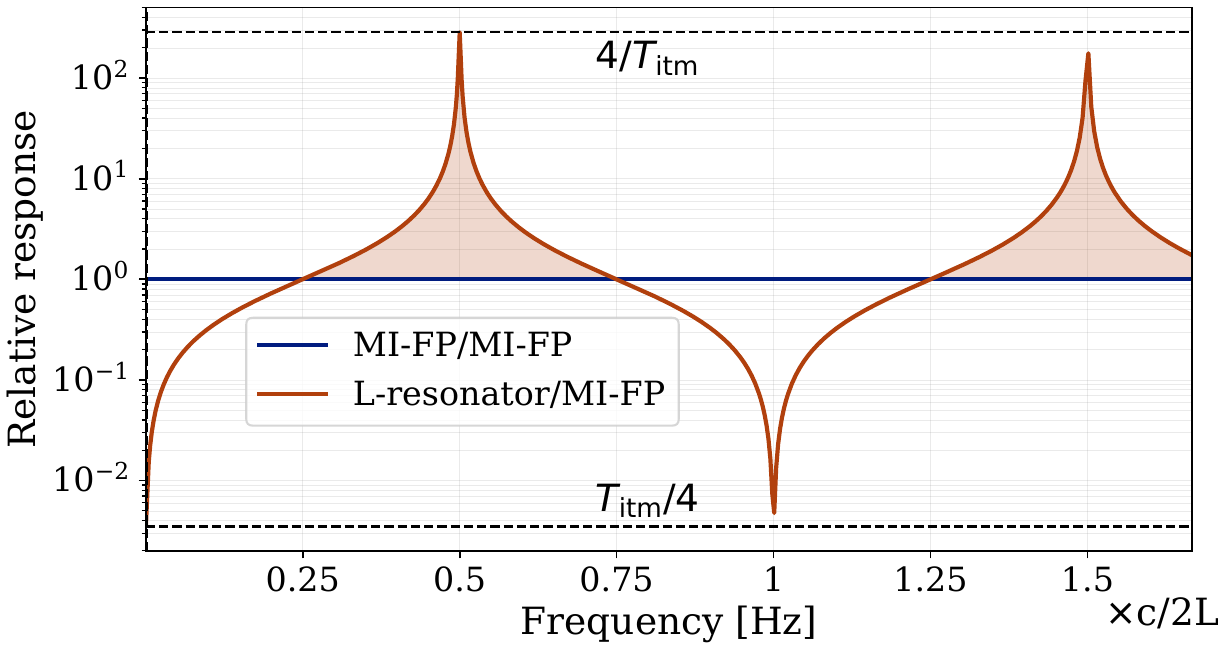}
     \caption{The relative response of the "L-resonator" and FPMI interferometer to the differential mode. The "L-resonator" outperforms at $c/4L$ and $3c/4L$ with a relative amplification of $4/T_{\rm itm}$, where $T_{\rm itm}$ is chosen as 0.014. In contrast, the FPMI interfeormeter outperforms around frequency 0\,Hz and $c/2L$.}\label{fig:Response}
\end{figure}

\section{Principle}\label{sec:Principle}
To illustrate the principle of the new scheme, we start with the fundamental FPMI interferometer. Here, we define two orthogonal degrees of freedom of motions: the common, ``$+$", and differential, ``$-$",  motions of end test masses (ETMs) in x- and y-arms,
\begin{equation}\label{eq:DL}
\Delta L_+=\Delta L_x + \Delta L_y\,, \Delta L_-=\Delta L_y - \Delta L_x\,.
\end{equation}
The differential mode signal transmits to the dark port, and the common mode signal appears at the bright port. The interferometer's responses to the sidebands of both modes are identical and proportional to
\begin{equation}
G_{\rm FPMI,+/-}=\frac{\sqrt{2T_{\rm itm}}}{\left|e^{2i\Omega \tau}- \sqrt{R_{\rm itm}}\right|}\,,
\end{equation}
where $\Omega$ is the angular frequency of sideband, $\tau\equiv L/c$, $L$ is the arm length,  $T_{\rm itm}$ and $R_{\rm itm}$ are the transmissivity and reflectivity of input test mass (ITM). 
\begin{figure}[t]
     \centering
     \includegraphics[width=0.73\linewidth]{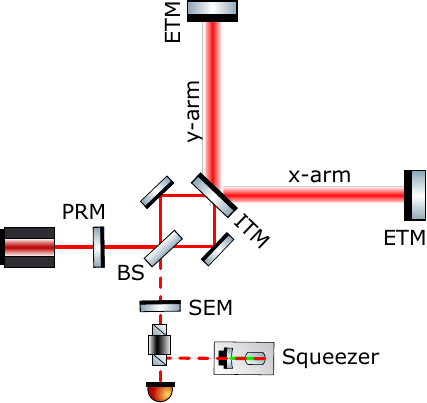}
     \caption{The schematic of new interferometer comprised of the ``L-resonator" and a central Michelson. It also includes power recycling mirror (PRM) and signal extraction mirror (SEM). Constant phase squeezing is injected from the dark port.}\label{fig:IFO}
\end{figure}

We propose an ``L-resonator" as shown in Fig.~\ref{fig:Cavity}. Such a resonator responds to the two modes of motion in different manners, giving separated resonant frequencies. The phase variation of light after a round trip in the cavity is
$\Delta \phi(t)=({2\omega_0}/{c} )[\Delta L_x(t-2\tau)+ \Delta L_y(t)]$, which, in the frequency domain, can be written as
\begin{equation}\label{eq:RTphi}
\Delta \phi(\Omega)=\frac{\omega_0}{c} (1 \pm e^{-2i\Omega\tau}) \Delta L_{\pm}(\Omega)\,.
\end{equation}
$\omega_0$ is the angular frequency of the carrier. 
It is clear that when $\Omega/2\pi=Nc/2L$ ($N$ is an integer), the phase shift of ``$+$" mode reaches maximum, in contrast, the ``$-$" mode peaks at $\Omega/2\pi=Nc/2L+c/4L$. 
Here we pump the resonator by two lasers from both input ports, which give balanced power in the two orthogonal arms\,\footnote{If the resonator is pumped from one port, the power at the two ETMs will be imbalanced and causes speed meter response for the differential mode\,\cite{korobko2020taming}.}. The signal of both modes of motion will appear at both ports as indicated in Fig.~\ref{fig:Cavity}. The responses of the two ports to the ``+" and ``-"  mode are proportional to
\begin{equation}\label{eq:RpL}
\begin{split}
    G_{L,1,+}=G_{L,2,+}=\frac{\sqrt{T_{\rm itm}}}{\left|e^{2i\Omega \tau}-  \sqrt{R_{\rm itm}}\right|}\,,\\
    G_{L,1,-}=-G_{L,2,-}=\frac{\sqrt{T_{\rm itm}}}{\left|e^{2i\Omega \tau}+ \sqrt{R_{\rm itm}}\right|}\,.
    \end{split}
\end{equation}
By combining and splitting the ``+" and ``-" mode signals at two ports, we can tell the ``L-resonator" has an identical ``+" mode response to the FPMI interferometer. Regarding to the ``-" mode, ``L-resonator" has $4/T_{\rm itm}$ times larger response at $c/4L$, as shown in Fig.~\ref{fig:Response}. The expense is around $\Omega=0$ and $c/2L$. It is worth noting that, at $\Omega=0$, the response is not zero, as one may infer directly from Eq.~\ref{eq:RTphi}.

\begin{figure}[t]
     \centering
     \includegraphics[width=1\linewidth]{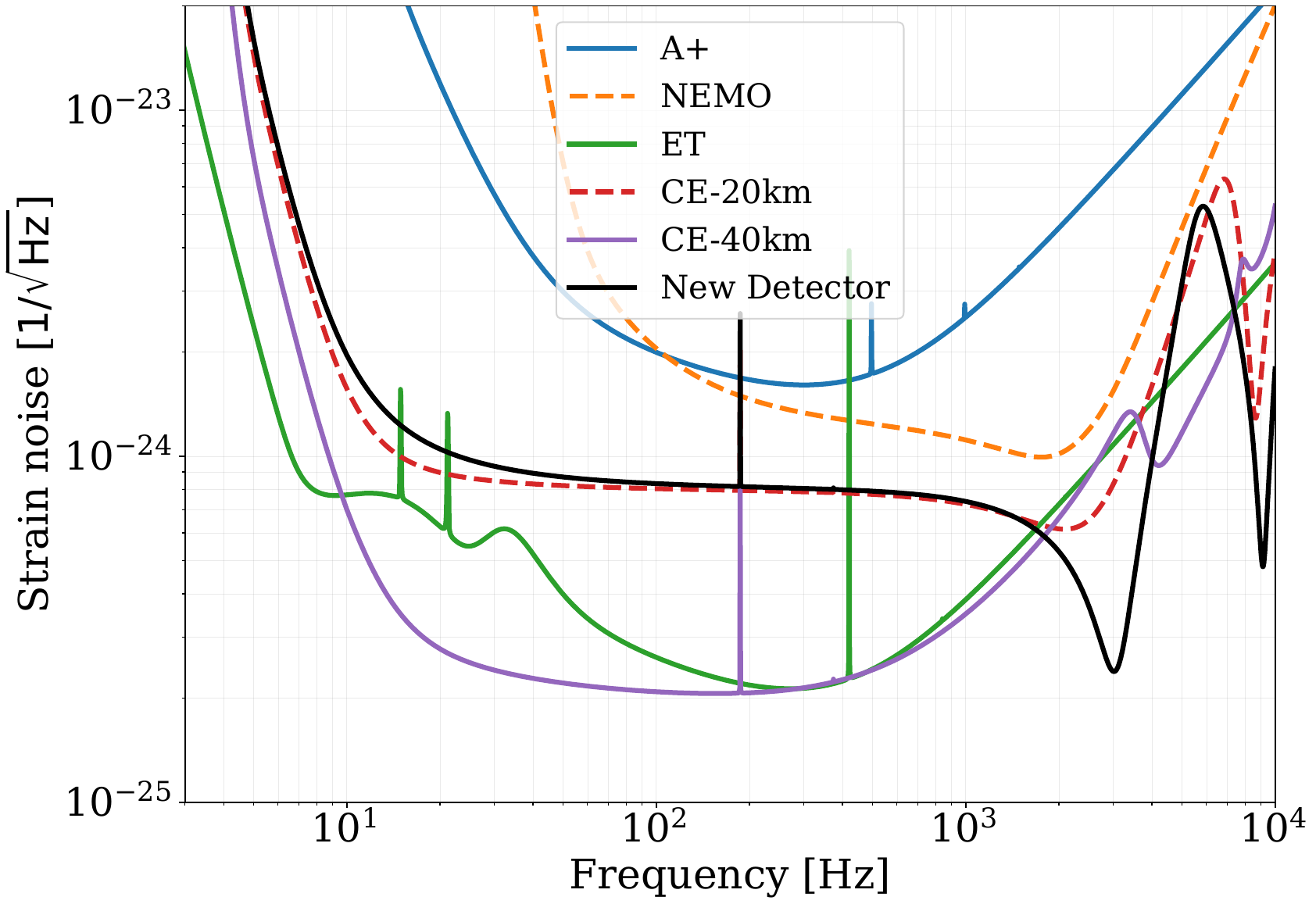}
     \caption{Strain sensitivity of the 25\,km new detector in comparison to A+\,\cite{2015,T1800042}, NEMO\,\cite{NEMO}, ET\,\cite{Hild_2011}, the ``post-merger" tuned CE-20\,km and the CE-40\,km \,\cite{evans2021horizon,Srivastava:2022slt} detectors. The sensitivity of ET corresponds to the combined three 10\,km triangular configurations.  The new detector gives superior sensitivity in the frequency range of 2-4\,kHz and 8-10\,kHz. The parameters of the new detector are listed in Table.~\ref{ta:par}.} 
     \label{fig:Sen}
\end{figure}

How about its response to gravitational waves?
Treating the ITM and ETM in x-arm/y-arm of the ``L-resonator" as the boundary of photon's trajectory in two directions, the round trip phase in one direction under the projection of gravitational waves is the same as in the Michelson interferometer where the beamsplitter (BS) and ETM form the boundary. The fractional change of travel time of light in one arm, $D(\Omega,n_e)$ can be expressed by Equation.~(5) in \cite{PhysRevD.96.084004}, where $n_e\equiv n_i e^i$. $n_i$ is the traveling direction of gravitational plane-wave, $e^{i}$ is the vector aligned to one arm. By calculating the optical field after traveling through the whole cavity, the cavity's response is proportional to $\sqrt{2}\omega_0\tau G_{L,1,-} D^{ij}$, where $D^{ij}=D(\Omega, n_k e^{k}_x) e^{i}_xe^{j}_x-D(\Omega, n_l e^{l}_y) e^{i}_ye^{j}_y$ is the detector tensor. It is the same as that of Michelson interferometer\cite{Rakhmanov_2008,Rakhmanov_2009,PhysRevD.96.084004,10.1093/acprof:oso/9780198570745.001.0001}. The antenna response for each polarization is $F_{+,\times}=D_{ij}\epsilon^{ij}_{+,\times}$, where $\epsilon^{ij}_{+,\times}$ is the polarization tensor for the $+$ and $\times$ polarizations\cite{PhysRevD.96.084004,PhysRevD.63.042003}, respectively. In the special case, the strain of normal incident waves can be mapped to an equivalent differential displacement as follows
\begin{equation}
\Delta L_-(\Omega)=L \frac{\sin\Omega\tau }{\Omega \tau}h(\Omega)\,.
\end{equation} It is clear that only at multiples of $c/2L$, instead of $c/4L$, the gravitational wave corresponds to 0 effective displacements.  In later sections, the strain sensitivities are shown by including the antenna response $\sqrt{|F_+|^2 + |F_\times|^2}$ for gravitational waves traveling at an angle of $15^{\circ}$ with respect to the normal direction.

\section{The complete interferometer and quantum noise}\label{sec:Quantum noise}
Each output port of the ``L-resonator" is
sensitive to both common and differential modes. We can decouple the two modes of signals through an electronic system after measuring their signals. More practically, we can use a single laser and add a BS to form the complete interferometer as shown in Fig.~\ref{fig:IFO}, which decouples the two modes spatially.
In Fig.~\ref{fig:IFO}, the sideband extraction mirror (SEM) and power recycling mirror (PRM) are also included. Note that in the lossless case, the laser travels in and returns along the same input port, and the reflection port turns out to be dark. Therefore, the BS behaves more like the one in the Michelson instead of the Sagnac interferometer, as the topology might indicate. The SEM helps decrease the sidebands' storage time around $c/4L$, therefore broadening the detector bandwidth\,\cite{MIZUNO1993273}. In contrast, it increases the low-frequency sideband storage time, which refers to the signal recycling scheme in DRFPMI interferometer\,\cite{MIZUNO1993273, mcclelland1995overview}.
A similar high-frequency resonance can be achieved by the synchronous interferometer\,\cite{drever1983interferometric, PhysRevD.38.433,PhysRevD.38.2317,PhysRevLett.101.101101,PhysRevD.77.022002}, which uses a ring cavity as its core. It realizes a speed-type measurement that suppresses the signals at low frequencies. The new interferometer is a position meter sensitive to signals toward DC.

The quantum noise of the new interferometer can be derived following the conventional approach of modeling the field propagation in the interferometer. The single-sided quantum noise power spectral density of the interferometer in the unit of $\rm m^2/ Hz$ is 
\begin{equation}
S(\Omega)=\frac{c^2\hbar|e^{2i\Omega\tau}+\sqrt{R_{\rm sec}}|^2}{4\omega_0 P_{\rm arm}T_{\rm sec}}+\frac{16\hbar \omega_0P_{\rm arm}T_{\rm sec}}{c^2M^2|e^{2i\Omega\tau}+\sqrt{R_{\rm sec}}|^2\Omega^4}\,,
\end{equation}
$M$ is the mass of each ETM.
Here $T_{\rm sec}$ and $R_{\rm sec}$ are the effective transmissivity and reflectivity of the SEC formed by ITM and SEM: 
\begin{equation}\label{eq:Tsec}
T_{\rm sec}\equiv \frac{T_{\rm itm}T_{\rm sem}}{\left[1-\sqrt{R_{\rm itm}R_{\rm sem}}\right]^2}\,,
\end{equation}
where $T_{\rm sem}, R_{\rm sem}$ are the power transmissivity and reflectivity of SEM. More precisely, we include the additional phase gained by the sidebands propagating in the SEC. There is
\begin{equation}\label{eq:QN}
S(\Omega)=\frac{c^2\hbar \left|\mathcal{C}\right|^2}{4\omega_0P_{\rm arm} T_{\rm itm}T_{\rm sem}}+\frac{16\hbar \omega_0 P_{\rm arm}T_{\rm itm}T_{\rm sem}}{c^2M^2 \left|\mathcal{C}\right|^2 \Omega^4}\,,
\end{equation}
where 
\begin{equation}
\mathcal{C}=e^{2i\Omega(\tau+\tau_s)}+\sqrt{R_{\rm itm}}\left[e^{2i\Omega\tau_s}-e^{2i\Omega\tau}\sqrt{R_{\rm sem}}\right]-\sqrt{R_{\rm sem}}\,.
\end{equation}
Here $\tau_s\equiv L_{\rm sec}/c$, $L_{\rm sec}$ is the SEC length. The first term of Eq.~\eqref{eq:QN} is the shot noise, $S_{\rm shot}$, and the second term denotes the radiation pressure noise.
The power spectral density of the SEC loss can be calculated as
\begin{equation}
S_{\rm sec}(\Omega)=\epsilon_{\rm sec}\left[\frac{c^2\hbar|e^{2i\Omega\tau}+\sqrt{R_{\rm itm}}|^2}{4\omega_0 P_{\rm arm}T_{\rm itm}}+ \frac{16\hbar \omega_0 P_{\rm arm}T_{\rm itm}}{c^2M^2 \left|\mathcal{C}\right|^2 \Omega^4}\right]\,
\end{equation}
where $\epsilon_{\rm sec}$ is the SEC loss coefficient. The power spectral density of the input loss is $S_{\rm in}(\Omega)=\epsilon_{\rm in}S(\Omega)$, where $\epsilon_{\rm in}$ is the input loss coefficient.
The power spectral density of the output loss is $S_{\rm out}(\Omega)=\epsilon_{\rm out} S_{\rm shot}(\Omega) /(1-\epsilon_{\rm out})$, where $\epsilon_{\rm out}$ is the output loss coefficient. The arm loss power spectral density can be calculated as
\begin{equation}
\begin{split}
S_{\rm arm}(\Omega)=\epsilon_{\rm arm}\left[\frac{c^2\hbar}{8\omega_0 P_{\rm arm}}
\right. \\
 \left. +\frac{8 \hbar\omega_0P_{\rm arm}|e^{2i\Omega\tau_s}-\sqrt{R_{\rm itm }R_{\rm sem}}|^2}{c^2M^2|\mathcal{C}|^2 \Omega^4}\right]\,,
\end{split}
\end{equation}
where $\epsilon_{\rm arm}$ is the arm loss coefficient.

\section{Conceptual design and noise budget}\label{sec:Conceptual}
Targeting the post-merger signals of binary neutron stars, typically between 2\,kHz and 4\,kHz, we propose a detector with each arm length 25\,km, which results in peak sensitivity at 3\,kHz. The sensitivities of the new detector, Advanced LIGO+ (A+)\,\cite{2015, T1800042}, and other third-generation detectors, Neutron Star Extreme Matter Observatory (NEMO)\,\cite{NEMO} Einstein Telescope (ET)\,\cite{Hild_2011} and Cosmic Explorer (CE)\,\cite{evans2021horizon,Srivastava:2022slt} are compared in the Fig.~\ref{fig:Sen}. 
\begin{figure}[t]
     \centering
     \includegraphics[width=1\linewidth]{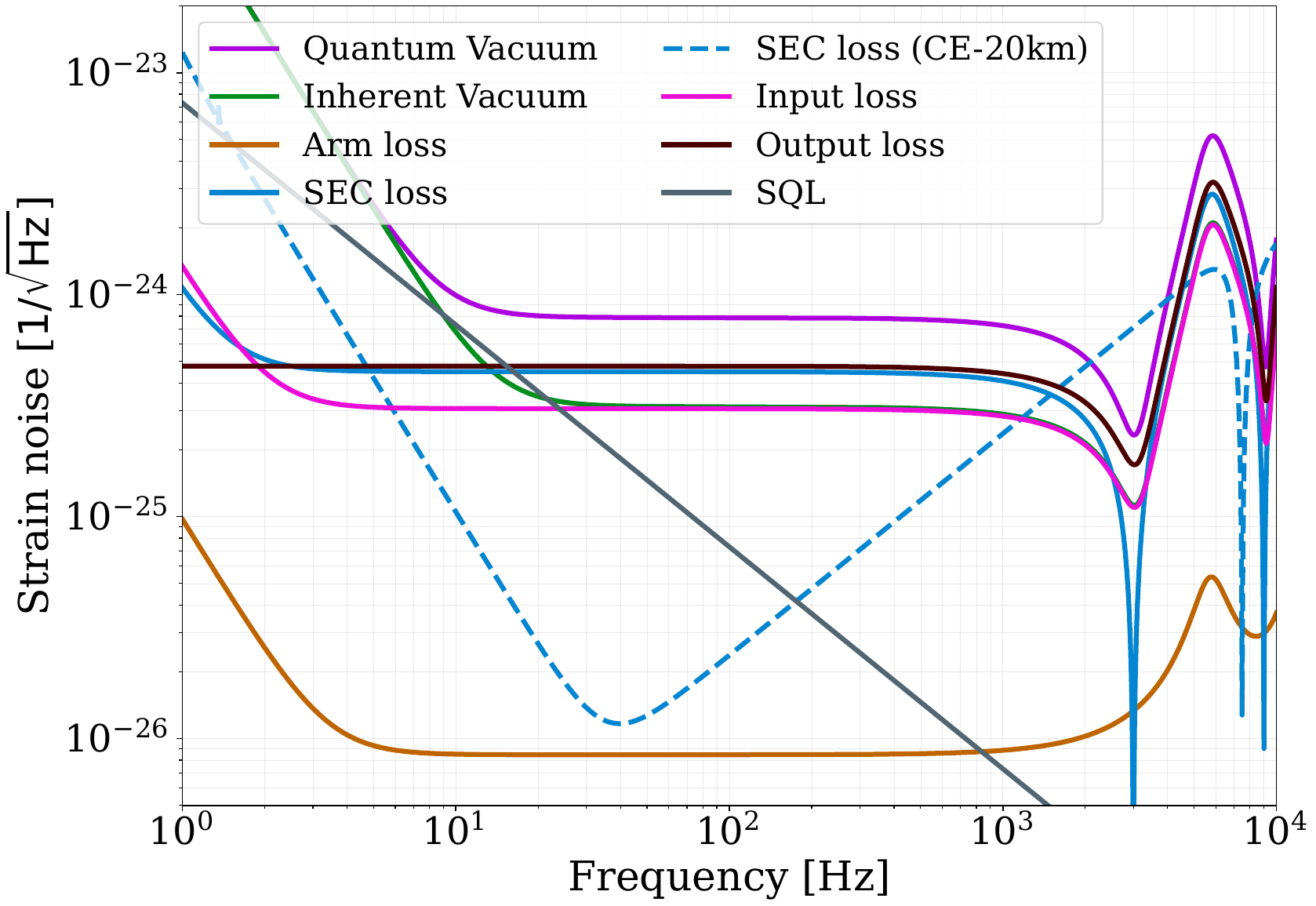}
     \caption{Detailed quantum noise budget of the new detector (solid lines). The SEC loss limit of the ``post-merger" tuned CE-20\,km detector (dashed line) is included for comparison. At 3\,kHz, the SEC loss limit of the new detector is orders of magnitude lower than that of the ``post-merger" tuned CE-20\,km detector. The output loss limits the quantum noise of the new detector. }\label{fig:Quantum}
\end{figure}
\subsection{Interferometer design}
In such a detector, we choose laser wavelength at 1064\,nm, and fused silica as the mirror material, same as in LIGO and Virgo detetcors\,\cite{2015,Acernese_2014}. We target the arm cavity power, 1.5\,MW, which can be obtained with 165\,W input laser and 100\,ppm round trip arm cavity loss (equivalent to 50\,ppm loss in each arm cavity of the DRFPMI interferometer). On the conceptual level, we reasonably assume other power degradation mechanisms, \textit{e.g.} the point absorbers on the surface of optics\,\cite{Brooks:21}, the parametric acousto-optic coupling\,\cite{PhysRevLett.114.161102} will be manageable in the future benefiting from both optimizing detector's technical design and technology research and development. The ITM transmissivity is chosen as 0.014. The SEM transmissivity is chosen as 0.06, and the SEC length is chosen as 50\,m. The resulting detector bandwidth is $\sim 440$\,Hz.
\begin{figure}[t]    
     \centering
     \includegraphics[width=1\linewidth]{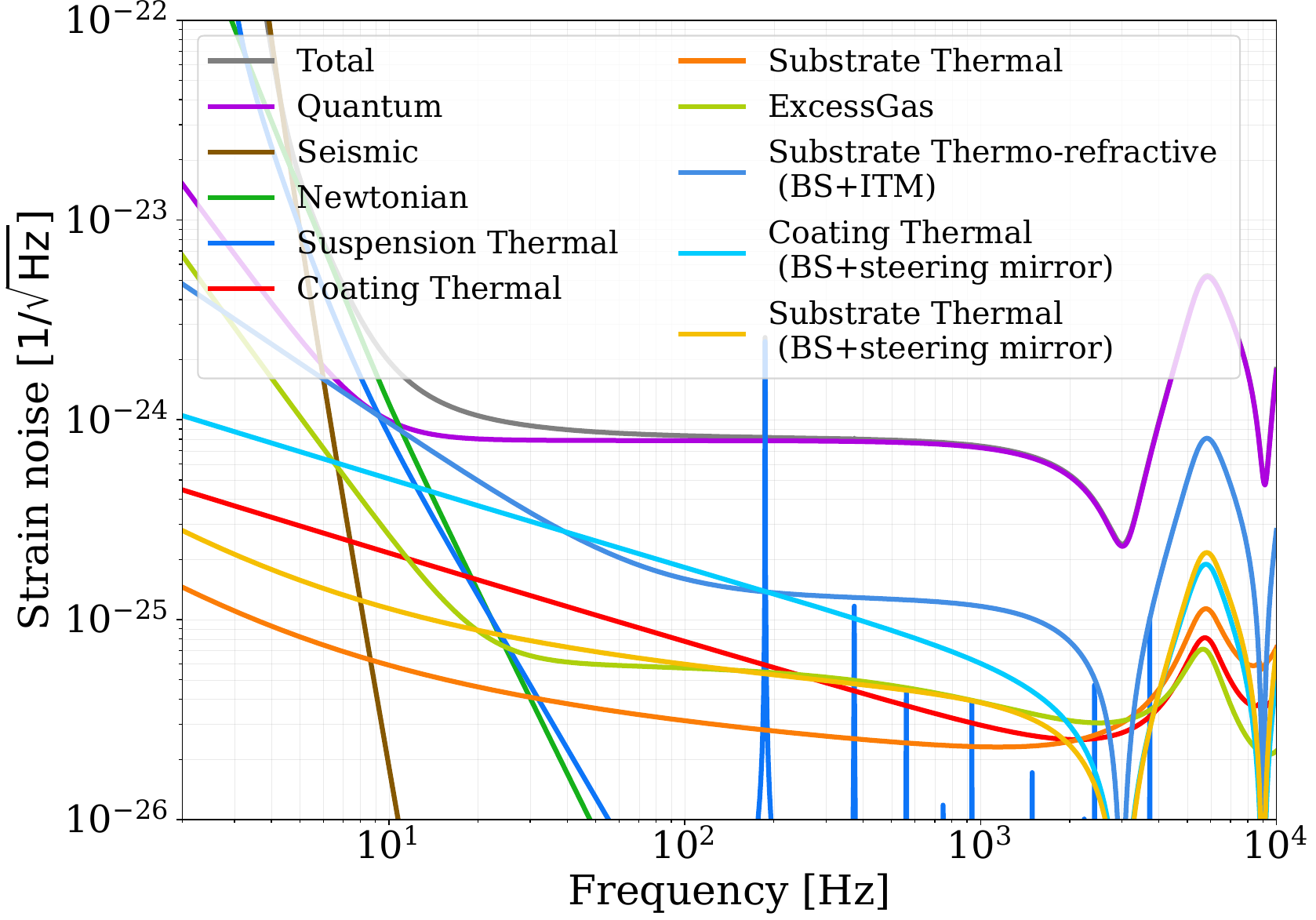}
     \caption{Noise budget of the 25\,km new detector. Above 20\,Hz, the sensitivity of the new detector is limited by quantum shot noise. Instead of frequency-dependent squeezing, only constant phase squeezing is applied. The seismic noise. Newtonian noise and suspension thermal noise include both arm cavity and central Michelson noise. We present the BS and ITM coating and substrate noise separately. The noise budget is modelled by the software, pyGWINC\,\cite{rollins2020pygwinc}.}\label{fig:NB}
\end{figure}

The radius of curvature of two ETMs is chosen to be 40\,km (cavity g factor $\sim 0.06$), resulting beam size of 13.2\,cm on ETMs. The beam is elliptical on the ITM, with sizes of 8\,cm and 11.3\,cm in two axes (11.3\,cm=$\sqrt{2}\times8$\,cm, where 8\,cm is the waist size). We adopt the mirror radius and thickness of 34\,cm and 40\,cm, respectively, giving total weight $\sim$ 320\,kg. In the new detector, the displacement of ITM along its normal is in the common mode, the sub-optimal geometry of the ITM, which leads to larger mirror thermal fluctuations\,\cite{PhysRevD.62.122002} is not a constrain from the noise perspective. We choose the ITM radius, 34\,cm, and thickness, 17\,cm, giving a mirror mass 136\,kg.  The coating material of ITM should have low optical absorption to reduce the heat load, hence reducing the optical loss in the recycling cavity from the thermal distortion\,\cite{Brooks:16}. However, it is not required to have a low mechanical loss, which is necessary to have by the ITMs in the DRFPMI interferometer. 

\subsection{Radiation pressure noise and squeezing}
In the new detector, only two ETMs contribute to the effective differential mode, which is half reduced compared with the DRMIFP interferometer. 
Equivalently, the reduced mirror mass of the differential mode is a factor of 2 larger; hence the quantum radiation pressure amplitude noise is reduced by a factor of 2. It is straightforward to read that Eq.~\eqref{eq:QN} reaches the minimum value when the shot noise and the radiation pressure noise are equal: 
\begin{equation}
S(\Omega)\geq\frac{4\hbar}{M\Omega^2}\,,
\end{equation}
which is the power spectral density of the standard quantum limit of the new interferometer. It is only half of the standard quantum limit of a DRMIFP interferometer\,\cite{kimble2001}. 

Benefiting from the naturally reduced radiation pressure noise, we only adopt constant phase squeezing instead of frequency-dependent squeezing. As a result, the usual kilo-meter-long filter cavity is not required. We assume 18\,dB constant phase squeezing can be generated. The loss projection on the input path is 1.5\%, including 1\% from the optical parametric amplifier and 0.5\% from the Faraday isolator. Note that the input loss budget is less than the estimation in other third-generation detectors, where a filter cavity is required on the input chain. It will result in 15\,dB squeezing into the interferometer and give 15\,dB amplification of radiation pressure noise. The output path, including the Faraday isolator (0.5\%), output mode cleaner (2\%), and the photodiodes (1\%), contribute total 3.5\% loss\,\cite{PhysRevLett.123.231107}. Including the internal loss from SEC (500\,ppm) and arm cavity (100\,ppm), 10-11.5\,dB squeezed shot noise will be observed over the whole frequency band, with 11.5\,dB squeezing at the peak sensitivity. The detailed quantum noise compositions are shown in Fig.~\ref{fig:Quantum}.

\subsection{Arm cavity noise}
The arm cavity effective mirror displacement noises are reduced compared with the equivalent DRFPMI interferometer, again benefiting from the in-susceptibility to ITM motions.

We model the noise budget of the new detector using the software pyGWINC\,\cite{rollins2020pygwinc}. The noise modeling is based on the CE-40\,km design\,\cite{PhysRevD.103.122004}. For the coating thermal Brownian noise modeling, we assume a factor 4 improvement from the mechanical loss of $\rm Ta_2O_5$ and $\rm SiO_2$ bi-layer coatings, which is also the goal of A+\,\cite{T1800042}. The ETM suspension design is assumed to be the quadruple pendulum suspension used in LIGO\,\cite{Aston_2012,PhysRevD.103.122004}. The vertical seismic and suspension thermal noise has the least coupling due to the finite radius of curvature of the earth and is independent of the arm length in the unit of strain. They will scale down by $\sqrt{2}$ from CE-40\,km. The horizontal seismic and suspension thermal noise will increase by 1.13 ($40/25/\sqrt{2}$). Newtonian noise is also around a factor of 1.13 times the CE-40\,km noise. The residual gas noise caused by the stochastic disturbance of molecular species onto the laser's phase\,\cite{P940008} is not mirror displacement noise and will almost scale up by 1.6 (40/25). The gas damping noise, however, from the impinging of the gas particles onto test masses\,\cite{CAVALLERI20103365, PhysRevD.84.063007} will scale up by 1.13. 

\begin{table}
\caption{Parameters of the 25\,km new detector}\label{ta:par}
\begin{ruledtabular}
\begin{tabular}{ccc}
    Wavelength & 1064\,nm \\
    Arm length & 25\,km \\
    SEC length &50\,m\\
    Input laser power & 165\,W \\
    Power recycling cavity power & 10.6\,kW\\
    Arm circulating power &1.5\,MW  \\
    ITM transmissivity & 0.014  \\
    SRM transmissivity & 0.06\\
    Arm loss & 100\,ppm\\
    SEC loss & 500\,ppm\\
    Input loss & 1.5\%\\
    Output loss & 3.5\%\\ 
    Input squeezing level & 18\,dB \\
    Result squeezing (shot noise) & 10-11.5\,dB\\
    Result squeezing (radiation pressure noise) & -15\,dB \\
    Substrate material & Silica\\
    ITM/ETM mass & 136\,kg/320\,kg \\
    ITM radius/thickness & 34\,cm/17\,cm\\
    ETM radius/thickness & 34\,cm/40\,cm \\
    ITM/ETM RoC & $\infty$/ 40\,km \\
    ITM/ETM beam size & 8\,cm/ 13.2\,cm\\ 
    BS radius/thickness & 34\,cm/17\,cm\\
    BS beam size & 8\,cm\\
    Coating loss angle (ETM/steering mirror) &  9e-5/1.25e-5
    
\end{tabular}
\end{ruledtabular}
\end{table}

\subsection{Central Michelson noise}
Different from the DRFPMI interferometer, where the arm cavity buildup factor attenuates the central Michelson noise, the optical paths fluctuations in the central Michelson of the new detector are non-negligible at low frequency and around $c/2L$ due to the anti-resonance of the arm cavity in those frequency bands.  The power spectral density of the Michelson differential noise can be mapped to the equivalent ETM noise as 
\begin{equation}
S_{\rm etm}(\Omega)=S_{\rm MI}(\Omega)\frac{|e^{2i\Omega \tau}+\sqrt{R_{\rm itm}}|^2}{4}\,,
\end{equation}
where $S_{\rm MI}(\Omega)$ is the central Michelson noise and $S_{\rm etm}(\Omega)$ is thus the detection noise. Such fundamental noise can largely come from the change
of the refractive index of the BS and ITM substrate due to inhomogeneous temperature
fluctuations, so called substrate thermo-refractive noise\,\cite{BRAGINSKY2000303,PhysRevD.80.062004,harry_bodiya_desalvo_2012}. The central Michelson thermo-refractive noise from both BS and ITM substrates can be calculated by Equation (2) in \,\cite{PhysRevD.80.062004}, including elliptical beam corrections. The beam size on BS will be almost the same as on the ITM. We chose the BS geometry the same as the ITM with a radius 34\,cm and a thickness 17\,cm. 

We also include the suspension, coating, and substrate thermal noise from the BS and the two steering mirrors between the BS and the ITM. The folding mirror gives coherent mirror displacement noise twice in a round trip. It also introduces larger mirror thermal noise in each bounce than a straight reflection due to the interference fringe pattern\,\cite{PhysRevD.90.042001, Sanders_2017}. It is 50\% addition in power spectral density for coating Brownian noise\,\cite{PhysRevD.90.042001, Sanders_2017}, the dominated thermal noise. The steering mirror coatings have the same mechanical loss as the ETMs. For the seismic noise and seismic Newtonian noise of the central Michelson, the scale of the central Michelson should be smaller than the wavelength of the seismic waves; the motions of Michelson mirrors can be largely coherent as a certain mix of differential and common modes. For simplicity, we assume uncorrelated noise from each mirror. The seismic isolation systems of central Michelson mirrors are the same as these of the ETMs. The detailed noise budget of the new detector is shown in Fig.~\ref{fig:NB}.

As shown in Fig.~\ref{fig:Sen}, the sensitivity of the new detector is comparable to that of the ``post-merger" tuned CE-20 km in the hundreds of Hertz range. It starts to have better sensitivity from 1.5\,kHz and shows more than a factor of 2 improvement in the frequency range from 2.6\,kHz to 3.8\,kHz with the maximal improvement of 3.5 times occurring at 3\,kHz.

\section{science cases}\label{sec:Science}
The new configuration leads to a broadband sensitivity as shown in Fig.~\ref{fig:Sen}, hence such a detector will still be able to probe the astrophysics of compact binaries and cosmology similar to third-generation detectors\,(see e.g. \cite{Maggiore_2020}), though with modified signal-to-noise ratios (SNRs). In this section, we highlight two specific science cases that become accessible due to the unparalleled sensitivity at high frequencies offered by the new design.

\subsection{Neutron star post-merger}
The primary science target of this new detector design is the post-merger gravitational wave signal of coalescing binary neutron star systems. While the ringdown spectroscopy of binary black hole coalescences provides a novel platform for tests of General Relativity\,\cite{LIGOScientific:2019fpa,LIGOScientific:2020tif,LIGOScientific:2021sio,Berti:2018vdi}, the post-merger spectroscopy of binary neutron stars will likely shed light on our understanding of the equation of state of nuclear matter at high temperatures, complicated magnetohydrodynamical processes involving neutrino generation and transport, novel phase(s) of nuclear matter and the underlying mechanism of short gamma-ray bursts\,\cite{Bauswein:2011tp,Bauswein:2012ya,Hotokezaka:2013iia,Takami:2014zpa,Paschalidis:2016agf,Metzger:2007cd,Bucciantini:2011kx,Bauswein:2018bma,PhysRevLett.128.161102}.

\begin{figure}[t]
\centering\includegraphics[width=1\columnwidth]{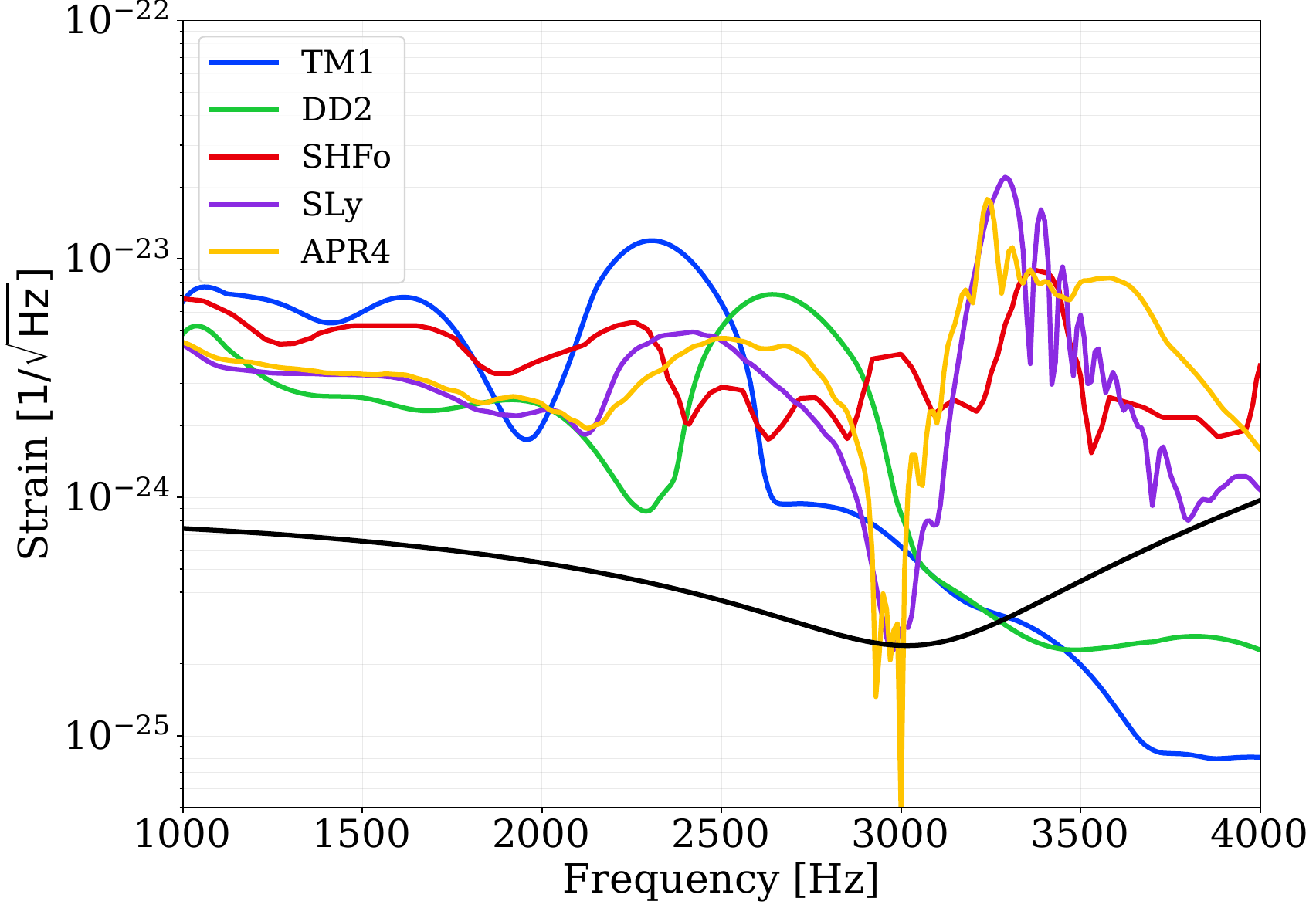}
\caption{The post-merger waveform of binary neutron stars in various of equations of states with peak mode frequency ranging from 2\,kHz to 4\,kHz. The source distance is assumed to be 100\,Mpc. Each of the neutron star in the binary is assumed to have equal mass, $1.35 M_\odot$. The black solid line represents the sensitivity of the new detector. } 
\label{fig:strain}
\end{figure}

The post-merger waveform of binary neutron stars is a complicated (and unknown) function of the neutron star masses, spins, and the equation of state. We perform an analysis similar to the one presented in \cite{PhysRevD.99.102004} and choose five representative equations of state: TM1~\cite{Sugahara:1993wz}, DD2~\cite{Typel:2009sy}, SFHo~\cite{Steiner:2012rk}, SLy~\cite{Reinhard:1995rf} and APR4~\cite{Akmal:1998cf} to cover a range of different stiffness. The neutron star binary is assumed to be $1.35 M_\odot+1.35 M_\odot$, and the corresponding waveforms are adopted from \cite{Takami:2014tva,Palenzuela:2015dqa,Sekiguchi:2011mc,Stergioulas:2011gd}. In Figure.~\ref{fig:strain}, we present the waveforms of a source in different equations of state at a distance of 100\,Mpc. Notice that the actual post-merger waveform depends on the remnant mass, spin, magnetic field level and possibly many other factors, in addition to equation of state. The goal of this new detector is to provide superior sensitivity covering from over 1\,kHz to 4\,kHz, where most post-merger spectra belong to. However, it is not necessary that the detector sensitivity is optimal for a particular waveform considering the uncertainty.

\begin{figure}[t]
\centering
\includegraphics[width=1\columnwidth]{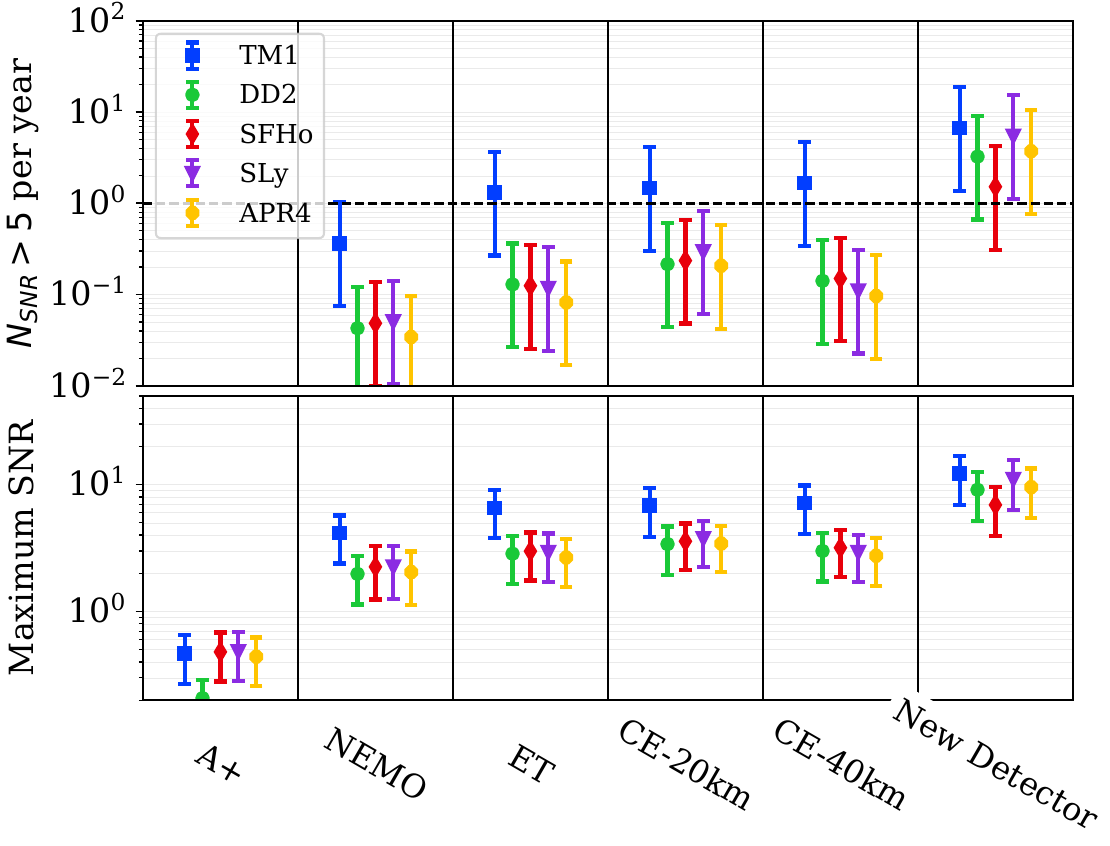}
\caption{ Number of events with SNR greater than five (top) and the SNR of the loudest event (bottom), both are assuming one year of observing time. They are obtained from the median values in the Monte-Carlo simulations.  The upper ends of the error bars correspond to the case with the merger rate being $295.7 {\rm Gpc}^{-3} {\rm yr}^{-1}$, the lower ends correspond to $21.6 {\rm Gpc}^{-3} {\rm yr}^{-1}$ and the symbols correspond to $105.5 {\rm Gpc}^{-3} {\rm yr}^{-1}$. } 
\label{fig:snrnevent}
\end{figure} 

We assume the updated binary neutron star merger rate after the third observing run of the Advanced LIGO-Virgo detector network, $105.5^{+190.2}_{-83.9} {\rm Gpc}^{-3} {\rm yr}^{-1}$~\cite{LIGOScientific:2021psn}. We note that it is an order of magnitude smaller than the initial rate estimated after the detection of GW170817 \cite{PhysRevLett.119.161101} (and used in \cite{PhysRevD.99.102004}). For each of the five equations of states, we apply Monte-Carlo simulations to randomly sample binary neutron stars in their sky locations, inclinations, and distances according to the merger rate, assuming one year of observation time and a uniform distribution of binaries in comoving volume.  We repeat this exercise  $100$ times to generate $100$ universes with the same underlying distribution but different statistical realizations. We compute the post-merger SNR for each event, defined as
\begin{align}
{\rm SNR} := 2\sqrt{\int^{4 {\rm k Hz}}_{f_{\rm contact}} df \frac{|\tilde{h}(f)|^2}{S_n(f)}}\,,
\end{align}
where $\tilde{h}$ is the frequency-domain post-merger waveform, $S_n(f)$ is the single-sided noise power spectral density of the considered detector, and $f_{\rm contact}$ is the frequency at neutron stars collide with each other \cite{Damour:2012yf,Takami:2014tva}, which depends on the equation of state. It can be computed from the mass ratio of the binary, and the compactness of the neutron stars \cite{Damour:2012yf}. The upper cut-off frequency is chosen to be 4\,kHz to cover most of the spectral power of the post-merger waveform.
We note that this is a slightly different definition of the post-merger SNR than the one used in ~\cite{Miao:2017qot,Srivastava:2022slt}, where instead of $f_{\rm contact}$ a fixed lower cut-off frequency of 1\,kHz is used. 
Here we choose the equation-of-state dependent contact frequency as we are not considering the part of the post-merger spectrum that overlaps with inspiral frequencies. The resulting median number of events with a post-merger SNR $>5$ and the median SNR of the loudest event, averaged over the $100$ realizations and assuming three different merger rates, are shown in Fig.~\ref{fig:snrnevent}. 
We find that the new design significantly outperforms -- by a factor of 1.7 to 4 -- other third-generation detectors in detecting the post-merger signal of colliding neutron stars. If the detection threshold is chosen to be SNR $>$ 5, then only the new design is confidently expected to observe at least one event per year, whereas the detection rates for NEMO, ET and CE are one to two orders of magnitude smaller for most of the equations of state. 
Notice that here the SNR is defined for the total post-merger waveform, which is greater than the SNR stemming from the excitation of individual oscillation modes of the post-merger object. Therefore, the new design provides the most promising platform to perform post-merger spectroscopy, i.e., resolving the ``peak mode" and other secondary oscillation modes of the remnant\,\cite{Yang:2017xlf}.

\subsection{Dark Matter Induced Neutron Star Collapse}
In addition to the prominent sensitivity between 2\,kHz and 4\,kHz, the new detector also has better sensitivity between 8\,kHz and 10\,kHz. This section studies its potential advantage for an even higher-frequency science case.

It has been proposed that fermionic dark matter particles may accumulate at the center of neutron stars and eventually collapse into a mini black hole because of the dissipative interaction. The neutron star matter subsequently accretes onto the mini black hole in millisecond timescale and produces gravitational wave emission in the kilohertz range, depending on the mass of the collapsing neutron star.

The primary gravitational wave signal for a rotating neutron star is the $20$ mode of the axisymmetric collapse. The collapse waveform in discussed in \cite{East:2019dxt}, which displays similar magnitude (in $\psi_4$) as the collapse waveform of the hypermassive neutron star as used is \cite{Zhang:2020qlh}. For simplicity, we shall use the phenomenological waveform model in \cite{Zhang:2020qlh} to compute the signal-to-noise ratio of these events:
\begin{align}
h =A \frac{50 {\rm Mpc}}{d} \sin (2 \pi f t)e^{-\pi f |t|/Q}
\end{align}
with the pre-peak part contributing comparable SNR as the post-peak part. Here the mode frequency is inversely proportional to mass $f \sim 4.7 {\rm kHz} \times (2.7 M_\odot/M_{\rm NS})$, the quality factor is $Q \sim 2.5$
and the amplitude is estimated as $A \sim 0.8 \times 10^{-23}(4.7 {\rm kHz}/f)^2$ (assuming the ``TM1" equation of state). 
Notice that these values are expected to change for different neutron star spin and equations of state. For example, the amplitude may change by at least a factor of three according to the equation of states used in \cite{East:2019dxt}, and the variation of frequency and quality factor may be $\le 30\%$ or $\le 20\%$ respectively \cite{Zhang:2020qlh}.

\begin{figure}[t]
\centering
  \includegraphics[width=1\columnwidth]{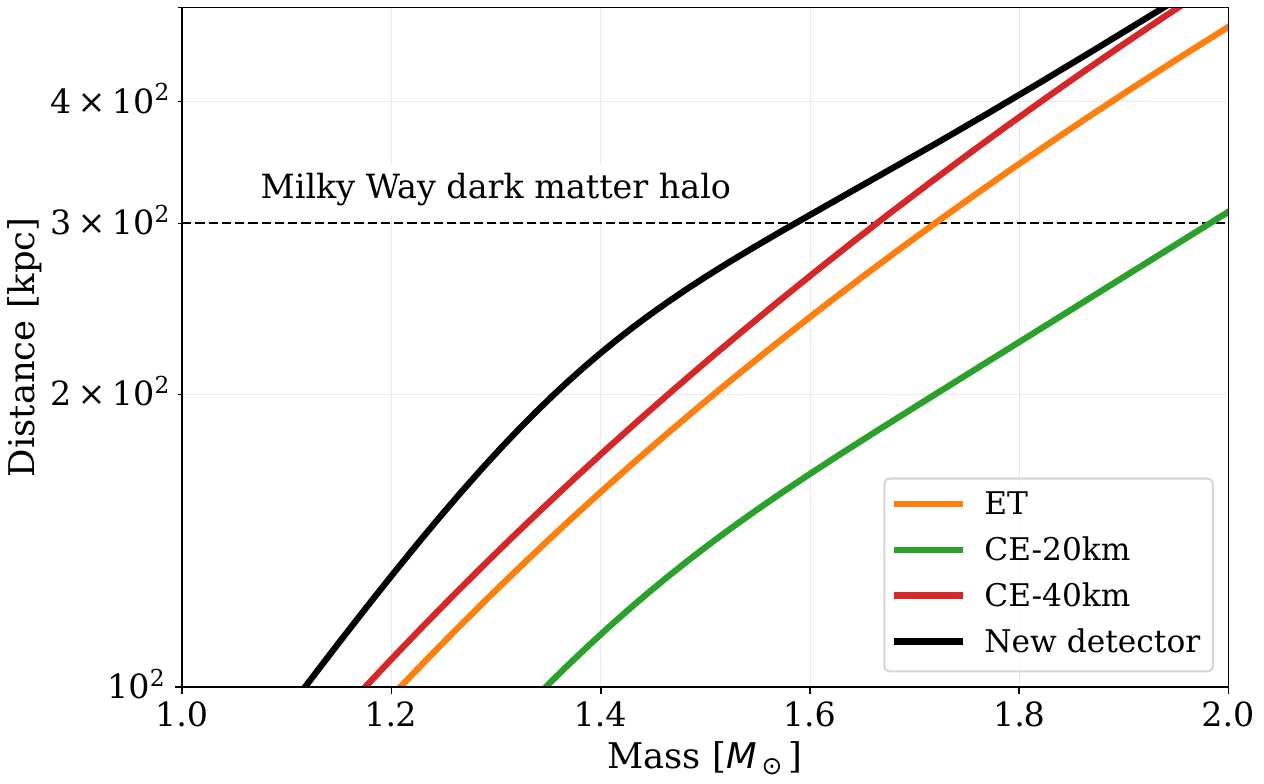}
\caption{ Horizon distance for detecting the dark-matter induced neutron star collapse signal with different detector sensitivities, assuming the threshold SNR  is 5. }
\label{fig:distance}
\end{figure}

As shown in Fig.~\ref{fig:distance}, with the waveform model of the collapse process, we can estimate the horizon distance if the detection threshold is set to be ${\rm SNR}=5$. The new design shows approximately 30\% improvements in horizon distance for neutron star mass around $1.4$ solar mass compared with other third-generation detectors, benefiting from the secondary dip in the sensitivity curve of the new detector in\,Fig.~\ref{fig:Sen}. The enhanced horizon distance is reaching toward the edge of Milky Way's dark matter halo for low mass collapse events.

\section{Conclusion}
As an extraordinary laboratory to study nuclear physics, the post-merger neutron stars call the demand to enhance the gravitational wave detectors' sensitivity in the kilohertz band. The currently existing and proposed gravitational wave detectors are based on the DRFPMI interferometer. Since the Fabry–P\'erot cavity is limited to having a narrow bandwidth to maintain its high finesse operation, such configurations have sub-optimal performance toward high frequencies. The fundamental limit from optical losses sets the barrier, particularly the loss in the SEC, which mixes with decayed high-frequency signals from the arm cavity. 

In this work, we present an elegant interferometer based on a ``L" shape optical resonator, which amplifies both the high-frequency gravitational wave signals and carrier. The SEC loss limited sensitivity of such an interferometer surpasses that of a DRFPMI interferometer by orders of magnitude at high frequencies, more precisely, $4/T_{\rm itm}$ at $c/4L$. Beyond its
high-frequency superiority, its in-susceptibility to ITM motions provides a factor of $\sqrt{2}$ suppression of displacement noises from the arm cavity mirrors and a factor of 2 suppression of the quantum radiation pressure noise. It also brings other advantages; for example, the constant phase squeezing turns out to be sufficient; the ITM coatings only require to satisfy low optical absorption without demanding low mechanical loss. The drawback of the new interferometer is at low frequencies around 0\,Hz, where the noises in the central Michelson couple to the detector readout with the same gain of arm cavity noise. On the technical level, the potential technical topics, \textit{e.g.} the parametric instability \,\cite{PhysRevLett.114.161102}, radiation-pressure-induced angular instability of the arm cavity\,\cite{Barsotti_2010} and required thermal compensation system\,\cite{Brooks:16} that will be different from the DRFPMI interferometer require separate study in the future.

We apply the sensitivity of the new detector to measure its ability to detect the binary neutron star post-mergers through the Monte-Carlo population simulation. Based on the chosen equation of states, the new detector has SNRs a factor of 1.7 to 4 of the other third-generation detectors. Taking the SNR $>$ 5 as the threshold and given the merger rate ranging from $21.6\,{\rm Gpc}^{-3} {\rm yr}^{-1}$ to $295.7\,{\rm Gpc}^{-3} {\rm yr}^{-1}$ according to GWTC-3 catalog, the new detector can enable a maximal detection rate upto $\sim $20 per year.

In addition, we explored the potential advantage of the new detector in its ability to detect dark-matter-induced neutron star collapse. Benefiting from the conspicuous sensitivity between 8\,kHz and 10\,kHz, which results from the second resonance of the new interferometer at $3c/4L$, it gives  $\sim$30\,\% improvement for neutron stars with around 1.4 solar mass, in comparison to the horizon reach of other configurations. 

\section{Acknowledgements} 
We thank LVK collaboration, Kevin Kuns, Daniel Brown, Mikhail Korobko, Jennifer Wright, Stefan Danilishin, Stefan Hild and Sebastian Steinlechner for fruitful discussions. T. Z., D. M., P. S. and H. M. acknowledge the support of the Institute for Gravitational Wave Astronomy at the University of Birmingham, STFC Quantum Technology for Fundamental Physics scheme (Grant No. ST/T006609/1), and EPSRC New Horizon Scheme (Grant No. EP/V048872/1
) T. Z. acknowledges the support of department of Gravitational Waves and Fundamental Physics in Maastricht University and E-TEST project. H. M. is supported by State Key Laboratory of Low Dimensional Quantum Physics and the start-up fund from Tsinghua University. D. M. is supported by the 2021 Philip Leverhulme Prize. H. Y. is supported by the Natural Sciences and
Engineering Research Council of Canada and in part by
Perimeter Institute for Theoretical Physics. Research at
Perimeter Institute is supported in part by the Government
of Canada through the Department of Innovation, Science
and Economic Development Canada and by the Province of
Ontario through the Ministry of Colleges and Universities.
P.~S. acknowledges support from STFC Grant No. ST/V005677/1.

\bibliography{bibliography}
\end{document}